\documentclass[a4paper]{article}
\usepackage[T1]{fontenc}
\usepackage{graphicx}
\usepackage[numbers]{natbib}
\usepackage[
    linesnumbered      % adds numbers for each line
]{algorithm2e}
\makeatletter
\renewcommand{\algocf@captiontext}[2]{\AlCapNameSty{\AlCapNameFnt{}#2\endgraf}}%
\makeatother
\usepackage{float}  % to place snippets in text, not floats

\usepackage[
    dvipsnames,         % adds additional named colors
]{xcolor}
\usepackage{amsmath, amsfonts}
\usepackage{enumitem} % configurable paragraph/figure numbering and spacing

\usepackage{listings} % simple code listing
\usepackage{simplebnf}  % for BNF of DSL
\usepackage[shortrightarrow]{stmaryrd} % just for fig 4!

\usepackage{tikz}
\usetikzlibrary{calc,shapes,arrows,arrows.meta}
\definecolor{matplotlibblue}{HTML}{1f77b4}
\definecolor{matplotlibgreen}{HTML}{2ca02c}
\definecolor{matplotliborange}{HTML}{ff7f0e}

\tikzset{
level/.style = {
         rectangle, 
         minimum height=0.8cm, 
         },
model/.style = {
         rectangle, draw,
         minimum width=3cm, minimum height=0.8cm, 
         fill=gray!20, text=black, 
         align = center, 
         },
lang/.style = {
         rectangle, draw,
         minimum width=3cm, minimum height=0.8cm, 
         fill=matplotliborange!20, text=black, 
         align = center, 
         },
set/.style = {
         rectangle, draw,
         minimum width=3cm, minimum height=0.8cm, 
         fill=matplotlibblue!20, text=black, 
         align = center, 
         },
comment/.style = {
         rectangle callout, callout relative pointer={(-1,0)}, 
         callout pointer width=4pt,
         draw,
         minimum width=3cm, minimum height=0.8cm, 
         color=green!60!black, fill=green!5, text=black, 
         align = center, font=\footnotesize\sffamily
         },
m1/.style = {
        rounded corners=4pt
        },
m2/.style = {
        rounded corners=8pt
        },
m3/.style = {
        rounded corners=12pt
        }
}

\title{Novel models of computation from novel physical substrates: a bosonic example}
\author{
Sampreet~Kalita$^1$ \and
Benjamin~W.~Butler$^1$ \and
Susan~Stepney$^2$ \and
Viv~Kendon$^1$
}
\date{\small
$^1$ Department of Physics, University of Strathclyde, G4 0NG, UK \\
\texttt{\{sampreet.kalita,benjamin.butler,viv.kendon\}@strath.ac.uk} \\
$^2$ Department of Computer Science, University of York, YO10 5DD, UK \\
\texttt{susan.stepney@york.ac.uk}
}
\begin{document}

\maketitle              % typeset the header of the contribution
\begin{abstract}
Unconventional physical computing is producing many novel and exotic devices that can potentially be used in a computational mode.
Currently, these tend to be used to implement traditional models of computation,
such as boolean logic circuits, or neuromorphic approaches.
This runs the risk of failing to exploit the devices to their full potential.
Here we describe a methodology for deriving a model of computation and domain specific language more closely matched to a given physical device's capabilities,
and illustrate it with a case study of bosonic computing as implemented
by a physical multi-component interferometer.

%\keywords{bosonic computing  \and photonics \and metamodel \and DSL \and PyBos}
\end{abstract}
%
%
%=========================
\section{Introduction}

Information is physical \cite{Landauer1991}; computing is physical.
Classical digital computing is physical: the work is done by physical CPU and GPU chips.
However, programming these devices uses high level languages far removed in abstraction level from the physical processes happening inside the chips.
Unconventional physical computing uses novel physical devices, with the hope of better exploiting their intrinsic physical properties.

A typical path to using unconventional physical computing devices takes a top down approach, starting from an existing model of computation, such as boolean logic circuits or neuromorphic approaches, and designing physical systems to conform to the model.
For more exotic physical systems, however, this runs the risk of not exploiting the physical properties to their best advantage.
Here we instead take a `bottom up' approach, examining the physical systems and their capabilities, and abstracting a suitable model of computation that directly exploits these.

We do this by defining a methodology for developing such a model.
Stepney~\cite{Stepney-2024-NACO} provides a tutorial for building a physical computer given a model of computation 
(in that case, the Echo State Network Reservoir Computing  model).
That method is based on the structure of physical computing
first defined in \cite{Horsman2014}.
It starts from an upper computational layer,
and defining a lower physical layer.
Here, we use the same structure, but in the opposite direction,
starting with a lower physical computing layer,
and using it to define the upper computational layer.
The aim is to give a structured approach
to defining an appropriate metamodel, domain specific language (DSL), and programming framework (sec.\ref{sec:metamodel}). 

We illustrate use of the method (outlined in sec.\ref{sec:method}, detailed in sec.\ref{sec:method:example}) through the development of a metamodel and DSL for bosonic computing as implemented by a physical multi-component interferometer.
%This is part of the LoCoMo (Lossy Computational Models) project.
Although the illustration is for a specific physical system,
our aim is for this method to be general enough that others can use it to derive models of computation suitable for their own physical substrates.

%===================================
\section{Background for the methodology: metamodelling for physical computing}\label{sec:metamodel}

Modelling and metamodelling are used in some branches of Software Engineering to ensure that descriptions of applications are formally defined, 
to enable them to be analysed, reasoned about, and transformed.
Here we take these ideas, and generalise them (slightly)
to provide a formal basis for defining our  methodology,
via an example metamodel and DSL (sec.\ref{sec:method:example}).

%===================================
\subsection{Classical approach}

\begin{figure}[tp]
\centering
\scalebox{0.85}{
\begin{tikzpicture}[font=\sffamily]

\node[level] (M0) at (0,0) {M0};
\node[model] (Mod0) at ($(M0)+(3,0)$) {System};
\node[comment] (m1comment) at ($(Mod0)+(4,0)$) {e.g. executing code,\\running interferometer};

\node[level] (M1) at ($(M0)+(0,2)$) {M1};
\node[model,m1] (Mod1) at ($(Mod0)+(0,2)$) {Model};
\draw[->,dashed] (Mod0) -- node[left] {instance} (Mod1);
\node[comment] (m1comment) at ($(Mod1)+(4,0)$) {e.g. set of equations,\\program code};

\node[level] (M2) at ($(M1)+(0,2)$) {M2};
\node[model,m2] (Mod2) at ($(Mod1)+(0,2)$) {Metamodel};
\node[lang,m2] (Lang2) at ($(Mod2)+(5,0)$) {Language};
\draw[->] (Mod2) -- node[above] {defines} (Lang2);
\draw[->,dashed] (Mod1) -- node[left] {instance} (Mod2);
\draw[->,dashed] (Mod1) -- node[right] {~written in} (Lang2);
\node[comment] (m2comment) at ($(Lang2)+(4,0)$) {e.g. UML, calculus,\\Python, DSL};

\node[level] (M3) at ($(M2)+(0,2)$) {M3};
\node[model,m3] (Mod3) at ($(Mod2)+(0,2)$) {Metametamodel};
\node[lang,m3] (Lang3) at ($(Lang2)+(0,2)$) {Metalanguage};
\draw[->] (Mod3) -- node[above] {defines} (Lang3);
\draw[->,dashed] (Mod2) -- node[left] {instance} (Mod3);
\draw[->,dashed] (Mod2) -- node[right] {~written in} (Lang3);
\node[comment] (m3comment) at ($(Lang3)+(4,0)$) {e.g. MOF, EBNF,\\category theory};

\end{tikzpicture}
}
\caption{\small The Software Engineering 4-layer modelling architecture.
The dashed lines represent relations between layers;
the solid lines represent relationships within layers.
See text for details.
}
\label{fig:kleppe}
\end{figure}
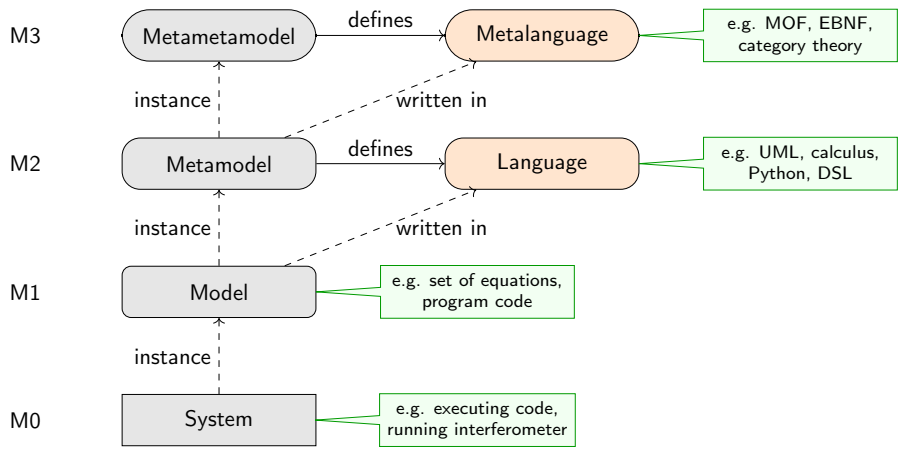

Kleppe et al \cite{Kleppe2003} outline the four-layer architecture and relationships between models and languages we use here
(see fig.\ref{fig:kleppe}).
The items in each layer are \textit{instances} of the items in the layer above.

\begin{enumerate}[label={M}\arabic*]\addtocounter{enumi}{-1} %start at 0
    \item \textbf{System or Instance layer}. This layer is the running system (physical or simulated) where the actual instances exist. Example: a given interferometer with a given arrangement of beam splitters and phase shifters, and a given input of photons. It is a particular execution or `run' of a real world experiment or computer simulation.
    \item \textbf{Model or Program layer}. This layer is the model of a set of related M0 systems, e.g. a mathematical model of a physical system, a computational specification of a simulation, or program code of a simulation.
    Example: a model of a generic interferometer, with parameters describing interferometer setup and inputs, which are instantiated for each layer M0 run.
    The model defines the concepts used to \textit{describe} its systems;
    a given system is an \textit{instance} of the model.
    \item \textbf{Metamodel or Programming Language layer}. This layer is the model of a set of related M1 models.    
    The metamodel defines the concepts used to \textit{describe} the  set of its models;
    a given model is an \textit{instance} of the metamodel.
    The metamodel \textit{defines} a modelling language or programming language; in some cases it might even be identified with the language it defines,
    but one metamodel may define many related languages; for example, an object-oriented metamodel might define the languages UML, Python, and C++.
    An M1 model is \textit{written in} the defined M2 language.
    A \textit{well-defined} language has a formally defined syntax (grammar) and semantics (meaning) so that it can be implemented by a computer: reasoned about, analysed, and executed in some manner.
    \item \textbf{Metametamodel or Metaprograming Language layer}. In the same way that an M1 model is written in an M2 language, an M2 metamodel is written in an M3 metalanguage, defined by an M3 metametamodel.
    This layer is typically the top layer:
    rather than requiring meta$^N$models, typically the M3 layer is defined in terms of itself, breaking the potential infinite regress.
    (This is because a language's syntax and semantics 
    are specified using a metalanguage;
    since a metalanguage is itself a language,
    its syntax and semantics can be defined by a metalanguage,
    which in many cases may be itself.)
    An M3-layer metalanguage may be less well defined than an M2-layer language, depending on usage. For example, many mathematical languages  are explained in natural language, or by example. Formalising these so that they can be processed by a computer is often a non-trivial exercise.
\end{enumerate}
%\notebb{Question 1: re ``typically the M3 layer is defined in terms of itself, breaking the potential infinite regress''. Why is this possible/is this always possible? Why is it possible at the M3 layer but not the M2 layer, or why don't we have to go to an M4 layer?}
%\notess{response: note added to explain this  [please comment out q/response when happy with response]}
%\notebb{Question 2: re ``An M3-layer metalanguage may be less well defined than an M2-layer language''. Why/what does this mean? Does ``well defined'' here mean in terms of a syntax and semantics?}
%\notess{response: note added to explain this  [please comment out q/response when happy with response]}

%===================================
\subsection{Extension for Methodology Definition}

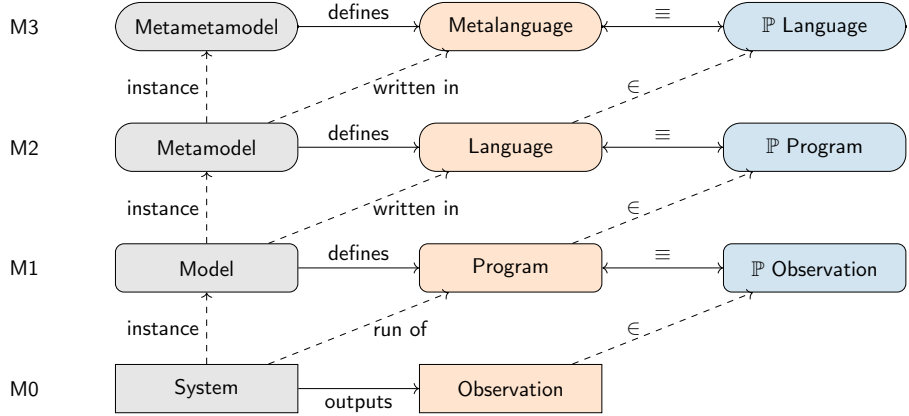
\begin{figure}[tp]
\centering
\scalebox{0.8}{
\begin{tikzpicture}[font=\sffamily]

\node[level] (M0) at (0,0) {M0};
\node[model] (Mod0) at ($(M0)+(3,0)$) {System};
\node[lang] (Lang0) at ($(Mod0)+(5,0)$) {Observation};
\draw[->] (Mod0) -- node[below] {outputs} (Lang0);

\node[level] (M1) at ($(M0)+(0,2)$) {M1};
\node[model,m1] (Mod1) at ($(Mod0)+(0,2)$) {Model};
\node[lang,m1] (Lang1) at ($(Lang0)+(0,2)$) {Program};
\node[set,m1] (Set1) at ($(Lang1)+(5,0)$) {$\mathbb{P}$ Observation};
\draw[->] (Mod1) -- node[above] {defines} (Lang1);
\draw[<->] (Lang1) -- node[above] {$\equiv$} (Set1);
\draw[->,dashed] (Mod0) -- node[left] {instance} (Mod1);
\draw[->,dashed] (Mod0) -- node[right] {~run of} (Lang1);
\draw[->,dashed] (Lang0) -- node[left] {$\in~~$} (Set1);

\node[level] (M2) at ($(M1)+(0,2)$) {M2};
\node[model,m2] (Mod2) at ($(Mod1)+(0,2)$) {Metamodel};
\node[lang,m2] (Lang2) at ($(Lang1)+(0,2)$) {Language};
\node[set,m2] (Set2) at ($(Set1)+(0,2)$) {$\mathbb{P}$ Program};
\draw[->] (Mod2) -- node[above] {defines} (Lang2);
\draw[<->] (Lang2) -- node[above] {$\equiv$} (Set2);
\draw[->,dashed] (Mod1) -- node[left] {instance} (Mod2);
\draw[->,dashed] (Mod1) -- node[right] {~written in} (Lang2);
\draw[->,dashed] (Lang1) -- node[left] {$\in~~$} (Set2);

\node[level] (M3) at ($(M2)+(0,2)$) {M3};
\node[model,m3] (Mod3) at ($(Mod2)+(0,2)$) {Metametamodel};
\node[lang,m3] (Lang3) at ($(Lang2)+(0,2)$) {Metalanguage};
\node[set,m3] (Set3) at ($(Set2)+(0,2)$) {$\mathbb{P}$ Language};
\draw[->] (Mod3) -- node[above] {defines} (Lang3);
\draw[<->] (Lang3) -- node[above] {$\equiv$} (Set3);
\draw[->,dashed] (Mod2) -- node[left] {instance} (Mod3);
\draw[->,dashed] (Mod2) -- node[right] {~written in} (Lang3);
\draw[->,dashed] (Lang2) -- node[left] {$\in~~$} (Set3);

\end{tikzpicture}
}
\caption{\small The augmented Software Engineering 4-layer modelling architecture.
Same notation as Fig.~\ref{fig:kleppe} with $\mathbb{P}$ denoting power set in the alternative definitions in blue.
See text for details.
}
\label{fig:meta}
\end{figure}

We are interested in eventually defining an M2-layer DSL.
We start at level M0 (specific physical and simulation experiments) using M1-layer concepts (the physical model of bosons).
In order to do this, we augment the architecture above with
more detail at the M0 and M1 layers, and an equivalent formulation of the language components (see fig.\ref{fig:meta}).

%----------------------------------
\subsubsection{M0 and M1 layer languages}
In fig.\ref{fig:kleppe}, languages are defined only at the M2 and M3 layer.
It is convenient to be able to talk about analogous structures at the lower layers.

At layer M1, we say that a model defines a \textbf{Program}. 
A program can be viewed as a language that formalises the set of possible observations from system runs.
For example, this might be the set of input/output pairs
(observing just the end output of system from a given starting state),
or the set of execution traces (observing the events occurring in the system as it runs, from a given starting state).
A given program can clearly have different observations made:
often it is just the output, but it could include the amount of time or memory used, and on debugging the system trace might be observed.
Also, physical systems and their simulations might have different observations available in principle: 
a quantum system cannot readily be observed as it executes;
and a classical simulation of it might output probabilities, rather than a single sample.
Considering the program to be a language itself allows
different executions or runs of that to be considered equivalent or not (have the same observation or not).
%\notebb{Question 3: why do we need to consider the program as a language to say whether two observations are equivalent or not?}
%\notess{response: we don't \textit{need} to (I do say `can be viewed as'), but it helps here to make everything uniform.
%Also, when computer scientists say `language' they can mean something very simple, like the definition of the structure of a set of strings, for example. [please comment out q/response when happy with response]}

At M0, we say that a system run results in an \textbf{Observation}.
This is the particular observation that the specific run yields, drawn from the full set of observations equivalent to the program.
It is the result of the experimental run or simulation run.
As noted above, a metamodel may be identified with the language it defines;
analogously, a running system may be identified with the observation it results in.

%----------------------------------
\subsubsection{Intensional and Extensional language definitions}

There are two ways of formally defining sets: intensional and extensional.
An \textit{intensional} definition defines a set by the properties of its elements, for example, the set of cubes being $cube = \{n^3\,|\,n \in \mathbb{N}\}$.
Membership is tested by testing if the defining properties hold: $m \in cube \iff \exists\, n \in \mathbb{N}\,|\,m = n^3$.
An \textit{extensional} definition defines a set by listing its elements, for example, the set of cubes is $cube = \{1,8,27,64,125,\ldots\}$; 
membership is tested by inspection. 
%\notebb{Question 4: Should this be the set $cube$ again, rather than the set of composite numbers?}
%\notess{ :-)}
%\notess{response: Hangover from previous example, fixed  [please comment out q/response when happy with response]}

Languages have both an intensional and extensional definition.
The intensional one is most common:
it is the definition of the syntax and semantics of the language.
The extensional one is the set of all programs that conform to, or can be generated by, the language definition.
So we can say that a programming language is equivalent to the set of all programs in that language ($Language \equiv \mathbb{P} Program$,
where $\mathbb{P}$ denotes the power set).
%\notebb{Question 5: $\mathbb{P}$ signifies the power set?}
%\notess{response: yes, note added  [please comment out q/response when happy with response]}
Hence at layer M2, the metamodel, the language it defines, and the set of all programs in that language, are essentially equivalent constructs,
and a model is an instance of, or an element of,
the relevant construct.

We have added this extra extensional definition to the layers
to clarify the relationship between layers M0 and M1,
that are relevant for experimental observations.
At the M0 system layer,
we have individual items; all higher layers are sets of items of the layer below:
models are sets of systems, metamodels are sets of models, etc.
By adding the extensional definition of these sets,
it becomes clear what these sets are,
and so what the instances look like.

%===================================
\subsection{Inclusion of physical computing}\label{sec:ART}

Horsman et al \cite{Horsman2014} introduce a model of physical computing
that can be used to determine when a physical system is performing a computation,
as opposed to just `doing its thing'.
This involves constructing a `sufficiently commuting' square of operations.
Here we will assume the system is sufficiently commuting
(has been demonstrated to be computing),
and place the various items in the 4-layer architecture.

\begin{figure}[tp]
\centering
\scalebox{0.8}{
\begin{tikzpicture}[font=\sffamily]

\node[model] (ModP0) at (0,0) {$\mathbf{p}$};
\node[model] (ModP1) at ($(ModP0)+(6,0)$) {$\mathbf{p'}$};
\draw[-{Stealth[scale=1.3]}] (ModP0) -- node [above] {$\mathbf{\widehat{H}}$} (ModP1);

\node[model] (ModA0) at (0,2) {$m_\mathbf{p}$};
\node[model] (ModA1) at ($(ModP1)+(0,2)$) {$m_\mathbf{p'} =_\epsilon m_\mathbf{p}'$};
\draw[-{Stealth[scale=1.3]},dashed] (ModA0) -- node [above] {$C$} (ModA1);
\draw[-{Stealth[scale=1.3]}] (ModA0) -- node[left, near start] {$\widetilde{\mathcal{R}}$} (ModP0);
\draw[-{Stealth[scale=1.3]}] (ModP1) -- node[right, near start] {$\mathcal{R}$} (ModA1);

\end{tikzpicture}
}
\caption{\small The physical compute cycle, adapted from \cite{Horsman2014}.
}
\label{fig:compcycle}
\end{figure}
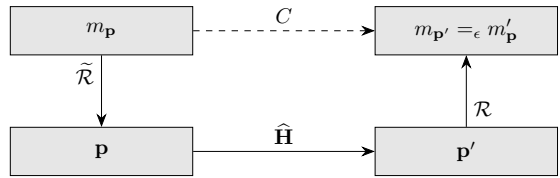

The model is shown in fig.\ref{fig:compcycle}.
The lower (physical) level shows a physical system evolving from an initial state $\mathbf{p}$ to a final state $\mathbf{p}'$ under physical process $\widehat{\mathbf{H}}$.
The upper (abstract) level shows an abstraction $m_\mathbf{p}$ of the physical state evolving to a final state $m_\mathbf{p}'$ under computational dynamics $C$.
The representation relation $\mathcal{R}$ links the physical and abstract levels.
The square $\epsilon$-commutes if the abstraction of the final physical state, $m_{\mathbf{p}'}$,
is sufficiently close to the final abstract state $m_{\mathbf{p}}'$, i.e., $m_{\mathbf{p}'}=_{\epsilon}m_{\mathbf{p}}'$.  In such a case,
we say that the physical system $\textbf{p}$ under the influence of physical process $\widehat{\mathbf{H}}$ performs the abstract computation $C$, given the representation $\mathcal{R}$.

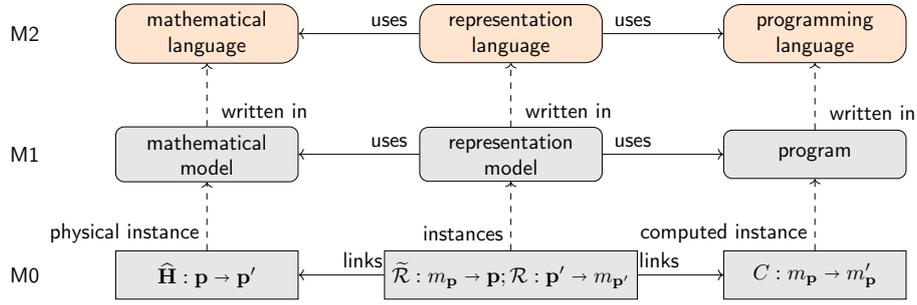
\begin{figure}[tp]
\centering
\scalebox{0.8}{
\begin{tikzpicture}[font=\sffamily]

\node[level] (M0) at (0,0) {M0};
\node[model] (ModP0) at ($(M0)+(3,0)$) {$\mathbf{\widehat{H}}: \mathbf{p} \shortrightarrow \mathbf{p}'$};
\node[model] (ModR0) at ($(ModP0)+(5,0)$) {$\widetilde{\mathcal{R}}:m_\mathbf{p}\shortrightarrow \mathbf{p} ; \mathcal{R}:\mathbf{p}'\shortrightarrow m_\mathbf{p'}$};
\node[model] (ModC0) at ($(ModR0)+(5,0)$) {$C: m_\mathbf{p} \shortrightarrow m_\mathbf{p}'$};
\draw[->] (ModR0) -- node [above, near start] {links} (ModP0);
\draw[->] (ModR0) -- node [above, near start] {links} (ModC0);

\node[level] (M1) at ($(M0)+(0,2)$) {M1};
\node[model,m1] (ModP1) at ($(ModP0)+(0,2)$) {mathematical\\model};
\node[model,m1] (ModR1) at ($(ModR0)+(0,2)$) {representation\\model};
\node[model,m1] (ModC1) at ($(ModC0)+(0,2)$) {program};
\draw[->] (ModR1) -- node [above, near start] {uses} (ModP1);
\draw[->] (ModR1) -- node [above, near start] {uses} (ModC1);
\draw[->,dashed] (ModP0) -- node[left, near start] {physical instance} (ModP1);
\draw[->,dashed] (ModR0) -- node[left, near start] {instances} (ModR1);
\draw[->,dashed] (ModC0) -- node[left, near start] {computed instance} (ModC1);

\node[level] (M2) at ($(M1)+(0,2)$) {M2};
\node[lang,m2] (LangP2) at ($(ModP1)+(0,2)$) {mathematical\\language};
\node[lang,m2] (LangR2) at ($(ModR1)+(0,2)$) {representation\\language};
\node[lang,m2] (LangC2) at ($(ModC1)+(0,2)$) {programming\\language};
\draw[->] (LangR2) -- node [above, near start] {uses} (LangP2);
\draw[->] (LangR2) -- node [above, near start] {uses} (LangC2);
\draw[->,dashed] (ModP1) -- node[right, near start] {~written in} (LangP2);
\draw[->,dashed] (ModR1) -- node[right, near start] {~written in} (LangR2);
\draw[->,dashed] (ModC1) -- node[right, near start] {~written in} (LangC2);

\end{tikzpicture}
}
\caption{\small The compute cycle components of fig~\ref{fig:compcycle} mapped to the 4-layer architecture
(see text for details).
}
\label{fig:compcyclelayers}
\end{figure}

We map these components onto the M0 layer of the 4-layer architecture
(fig.\ref{fig:compcyclelayers}).

\subsubsection{Layer M0 -- system instances}
\begin{itemize}
    \item     The physical instance $\textbf{p}$ is the initial state of a physical system. It evolves to final state $\textbf{p}'$ under the physical process $\widehat{\mathbf{H}}$.
    \item     The abstract instances $m_\textbf{p}$ and $m_\textbf{p}'$ are the initial and final abstract system states. They are linked by the semantics of the given abstract computation $C$.
    \item     The initial state instances $m_\textbf{p}$ and $\textbf{p}$ are linked  under the inverse representation relation $\widetilde{\mathcal{R}}$.
    The final state instances $\textbf{p}'$ and $m_{\textbf{p}'}$ are linked under the representation relation $\mathcal{R}$.
    Since we assume that the compute cycle has been verified to `sufficiently commute', we have $m_{\textbf{p}'} =_\epsilon m_{\textbf{p}}'$ (see fig.~\ref{fig:compcycle}).
\end{itemize}
\subsubsection{Layer M1 -- system models}
\begin{itemize}
    \item The behaviour of the specific physical system is captured by a mathematical model of the physical process in action.
    \item The behaviour of the specific abstract system
    captured by a program.
    \item The specific representational links between physical and abstract instances are captured in a representation model. This model will use parts of the physical dynamics model and the  abstract program in order to create the representation.
    \end{itemize}
\subsubsection{Layer M2 -- languages}
    (There may be separate M2 metamodels associated with these languages, but here we assume that these are the same.)
\begin{itemize}
    \item  The mathematical model of the physical dynamics is written in some mathematical language, defined at M2.
     \item  The program is written in some programming language, such as a DSL, defined at M2.
    \item The representation model is written in some representation language; this uses parts of both the mathematical language of the physical system,
    and the programming language of the computational system,
    to link the relevant concepts between the two.
\end{itemize}

%===================================
\subsection{Implementing a DSL for bosonic computing}\label{sec:impl_DSL}

Our aims include defining a new model of (lossy bosonic) computation, that is, a new metamodel.
Specifically, we do so by defining an
\textit{internal DSL} (see, for example, \cite{Fowler2011}), a programming language specific to the domain of bosonic computing,
that can be readily represented by the physical dynamics of photons and interferometers.
As an internal DSL, it is implemented inside an existing programming language,
here, Python.

%----------------------------------
\subsubsection{Using a simulation framework}
\begin{figure}[tp]
\centering
\scalebox{0.8}{
\begin{tikzpicture}[font=\sffamily]

\node[level] (M0) at (0,0) {M0};
\node[model] (ModC0) at ($(M0)+(3,0)$) {execution};

\node[level] (M1) at ($(M0)+(0,2)$) {M1};
\node[lang,m1] (LangDSL1) at ($(M1)+(3,0)$) {DSL program};
\node[lang,m1] (LangPy1) at ($(LangDSL1)+(5,0)$) {PyBos-Sampler fwk};
\draw[->] (LangDSL1) -- node [below, near start, xshift=1mm] {imports} (LangPy1);
\draw[->,dashed] (ModC0) -- node[left, near start] {instance} (LangDSL1);

\node[level] (M2) at ($(M1)+(0,2)$) {M2};
\node[lang,m2] (LangDSL2) at ($(M2)+(3,0)$) {DSL};
\node[lang,m2] (LangPy2) at ($(LangPy1)+(0,2)$) {Python language};
\draw[->,dashed] (LangDSL1) -- node[left, near start] {} (LangDSL2);
\draw[->,dashed] (LangDSL1) -- node[left, near start] {written in~~} (LangPy2);
\draw[->,dashed] (LangPy1) -- node[left, near start] {written in} (LangPy2);

\end{tikzpicture}
}
\caption{\small Implementing an internal DSL using a simulation framework.  DSL programs are written in a combination of the internal DSL and the host language, such as Python. 
The supporting framework is written in the host language,
to implement the various DSL constructs.
}
\label{fig:dslfwk}
\end{figure}
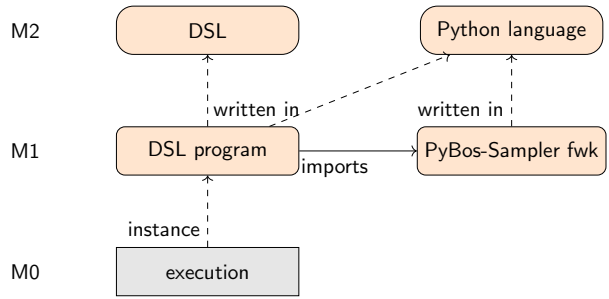

An internal DSL program can be implemented in simulation by using a \textit{framework} (fwk).
This involves developing a framework in a standard language (for example, our framework called PyBos-Sampler) that supports the DSL language concepts,
then developing a Python program using the framework to implement a Python representation of the DSL program
(see fig.\ref{fig:dslfwk}).
In some cases, the meaning of the DSL is only implicitly defined by its framework implementation;
in order to implement the same DSL in different programming languages,
it is necessary to extract a language-neutral definition to work from
\cite{Bock2025}.
For PyBos-Sampler, we have defined the semantics of the DSL separate from the framework implementation (see sec.\ref{sec:method:dsl}).

%----------------------------------
\subsubsection{Using domain-specific hardware}
\begin{figure}[tp]
\centering
\scalebox{0.8}{
\begin{tikzpicture}[font=\sffamily]

\node[level] (M0) at (0,0) {M0};
\node[model] (ModC0) at ($(M0)+(3,0)$) {sw execution};
\node[model] (ModIf0) at ($(ModC0)+(10,0)$) {hw device run};
\draw[->] (ModC0) -- node[below, near start] {calls} (ModIf0);

\node[level] (M1) at ($(M0)+(0,2)$) {M1};
\node[lang,m1] (LangDSL1) at ($(M1)+(3,0)$) {DSL program};
\node[lang,m1] (LangPy1) at ($(LangDSL1)+(5,0)$) {PyBos-Runner fwk};
\node[lang,m1] (LangIf1) at ($(LangPy1)+(5,0)$) {device program};
\draw[<-] (LangPy1) -- node [below, near end] {~~imports} (LangDSL1);
\draw[->] (LangPy1) -- node[below, near start] {calls} (LangIf1);
\draw[->,dashed] (ModC0) -- node[left, near start] {instance} (LangDSL1);
\draw[->,dashed] (ModIf0) -- node[left, near start] {instance} (LangIf1);

\node[level] (M2) at ($(M1)+(0,2)$) {M2};
\node[lang,m2] (LangDSL2) at ($(M2)+(3,0)$) {DSL};
\node[lang,m2] (LangPy2) at ($(LangPy1)+(0,2)$) {Python language};
\node[lang,m2] (LangIf2) at ($(LangIf1)+(0,2)$) {device language};
\draw[->,dashed] (LangDSL1) -- node[left, near start] {} (LangDSL2);
\draw[->,dashed] (LangDSL1) -- node[left, near start] {written in~~} (LangPy2);
\draw[->,dashed] (LangPy1) -- node[left, near start] {} (LangPy2);
\draw[->,dashed] (LangPy1) -- node[left, near start] {written in~~} (LangIf2);
\draw[->,dashed] (LangIf1) -- node[left, near start] {written in} (LangIf2);
(LangDSL2);

\end{tikzpicture}
}
\caption{\small Implementing an internal DSL using a hardware framework.  DSL programs are written in a combination of the internal DSL and the host language, such as Python. 
The supporting framework is written in a combination of the host language and the hardware device driver language,
which calls the hardware device to implement the various DSL constructs directly.
}
\label{fig:dsldirect}
\end{figure}
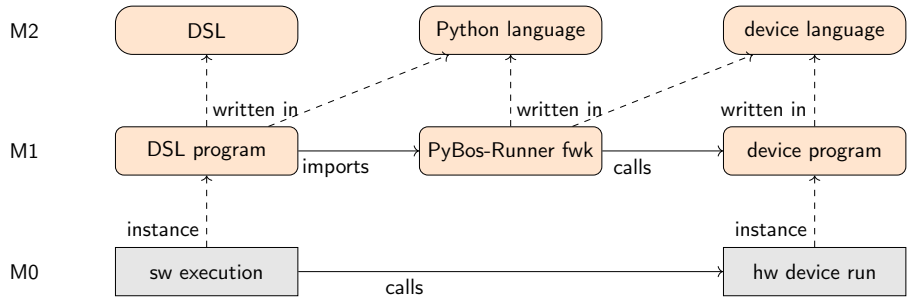

We wish to use special purpose hardware devices to gain computational advantage when executing bosonic computing programs.
A DSL framework supporting the device (such as our framework called PyBos-Runner, supporting the same DSL with the same interface as PyBos-Sampler) accesses the hardware (through a device program such as a device driver) rather than simulating the result in software
(see fig.\ref{fig:dsldirect}).
Pragmas (directives) may be used to allow the same framework to be used for simulation and hardware access,
supporting testing in simulation then more efficient runs on hardware from the same source code.
%The step from DSL program to hardware interferometer program is acting as the representation program, linking computational model to hardware implementation.

%==============================
\newpage
\section{Overview of the methodology}\label{sec:method}
In summary, our methodology for abstracting a metamodel and DSL is as follows:
\begin{enumerate}
    \item \textbf{Model the physical system.}  Develop a formal  model of the physical system, substrate and devices,
    sufficient to build a simulation.
    \item \textbf{Design, build and test a simulator} of the experimental setup.
    \item \textbf{Abstract out the metamodel concepts} from the model in computational terms.
    \item \textbf{Define a DSL}, both syntax and semantics, while respecting the physical system constraints.
    \item \textbf{Build a reference implementation} of the DSL.
    \item \textbf{Implement a representative sample of computational problems} using the DSL and reference implementation.
    \item \textbf{Implement the same problems using the physical device}.
\end{enumerate}
We flesh out these steps, and illustrate them with an example, in the following section.

In practice, the process progresses through several iterations of increasing sophistication
(as in the example below, starting with a lossless model, followed by a lossy one).
Each iteration can provide test cases and motivation for the next.

%------------------------------------
\newpage
\section{Example: from boson sampling physics to PyBos DSL}\label{sec:method:example}
Here we illustrate each step of the methodology in the context of computing with lossy interacting photonic systems.

\subsection{Model the physical system}\label{sec:eg:model}

Step 1 is to develop a formal (mathematical) model of the physical system, 
including both the physical substrate and the engineering used to manipulate the system.
This provides a model of the physical process $\mathbf{\widehat{H}}$ shown in the lower branch of fig.\ref{fig:compcycle}.

For devices corresponding to well-developed (advanced) text-book physics, no experimental investigations are needed to build the model.
In other systems, there may need to be further experimental work to characterise both the substrate and the way it can be configured or manipulated.
This might result in a phenomenological model, with parameters set by some machine learning exercise,
such as the neural ODE and SDE models from \cite{Manneschi-2025}.
If the devices are well-approximated by theory,
a hybrid approach may be used: a theoretical model with experimentally derived first order corrections.

\subsubsection{Example}
For our example, the domain is (lossy) photonic interferometers, where we analyse the photon statistics of their components and their output. A characteristic observation in such photonic hardware is photon `bunching' \cite{hong_measurement_1987}: if two indistinguishable photons enter a 50--50 beam splitter on different modes,
they will both exit on the same mode.
This potentially allows efficient implementation of the `boson sampling problem', which involves calculation of a matrix's \textit{permanent}, a computationally hard problem \cite{Aaronson2010,Aaronson2011}.

\paragraph{Basic model.}
We start with a basic model of a `perfect' system: lossless, with discrete photon modes.
The physics is the bosonic physics of interacting photons,
which is well-developed (advanced) text-book physics.
The basic \textit{physics model} is described in terms of the behaviour of simple optical devices: phase shifters and beam splitters, and how they transform input modes into output modes.  Modes contain photons, which are modelled using creation and annihilation operators.
The engineering is the bringing together of the photons to interact in an interferometer built from multiple components connected via different modes.
The theory of such interferometers is also well developed.
%The LoCoMo basic \textit{interferometer model}, which uses concepts from the physics model, is described in \S\ref{sec:sampling_interferometer}.

%A high level model of the operation of physical system is given as alg.\ref{algorithm:boson_sampling_experiment_vS&B}.
%A high level mathematical model of its behaviour is given as alg.\ref{algo:photodetection_maths}.

\paragraph{Augmented model.}
The basic model of a `perfect' system is  
augmented with two lossy gate models,
capturing more closely the systems that are implementable.
%\textbf{Loss model.}
%Two loss models are defined.
One lossy gate model allows photons to be lost in an uncorrelated manner during their passage through the interferometer.
This heuristic model of a lossy device is a composite system consisting of the ideal (lossless) device, with fictitious beam splitters placed before it that can lose input photons into new unobserved modes in an uncorrelated manner.
%\notess{just to check, \S\ref{sec: BS loss model} is LossOne, and \S\ref{sec: Noise op loss model} is LossTwo??}
%\notesk{Response: Yes, that's correct.}
The other lossy gate model assumes that the losses are internally correlated inside a single beam splitter.
This model also involves additional unobserved modes.

%\textbf{Gaussian modes model.} A continuum model, that moves from discrete photon modes to more realistic Gaussian wavepackets, is presented in chapter~\ref{ch:classical}.

%\notess{Please check/correct my terminology!}

\subsection{Design, build and test a simulator}\label{sec:method_build_sim}
The physical model of the system from step~1 forms the basis of developing the relevant metamodel and DSL.
As a route to abstracting out this model,
 step~2 is building a simulator,
with the aim of identifying relevant computational primitives and structures.

%The definitions from the model to develop a suitable design and pseudocode for the simulation.
%Such pseduocode for LoCoMo is presented in chapter~\ref{ch:sampling}, alg.\ref{algo:photodetection_simulation_vBen2.0}.

\paragraph{Basic simulator.}
The model is used to develop a design and pseudocode,
which is then refined into a basic simulator.
This can act as a testbed for correct implementation of the physical model,
by running test cases on small systems with known outputs.
If these known outputs are calculated directly from the model, any discrepancies are the fault of the implementation.
If these outputs are calculated independently of the model,
for example, being `well-known results',
then discrepancies might point to a problem with the model itself.

\paragraph{Optimisation.}
The basic implementation is likely to be inefficient.
The performance may be optimised, with the basic, tested version available as a `test oracle', providing known correct answers to test cases of the optimised simulator.
However, it is important that any such optimisation does not obscure the structure of the underlying system.
The purpose of building a simulator is to help abstract out a metamodel, not (merely) to have an efficient low-level simulation of the physical device.

\paragraph{Refactor to function calls.}
The basic simulator (or its optimised version)
should be refactored to ensure that basic operations are implemented uniformly as functions or procedures.
This allows the simulator to become the basis for the eventual DSL reference implementation.

\subsubsection{Example}
In our example system, the basic model is used to guide the design of a basic simulator written in Python.
Optimisation is done by using sparse algebra, and parallelising the implementation on multiple CPUs/GPUs using JAX \cite{jax2018github}.
% a Python library for high performance parallel computing on multiple CPU cores and multiple GPUs, with JIT (just in time) compilation.

%................................
\subsection{Abstract out the metamodel concepts}

The next step is to abstract out suitable metamodel concepts.
The physics and engineering models and simulator pseudocode should be interrogated for computational analogues to form the computational metamodel concepts.

The first metamodel for use with a novel physical device will most likely be at the level of a corresponding `assembly language', with metamodel concepts fairly directly related to physical properties and operations.
It will also likely draw on existing computational ideas.
Later abstractions can start from the computational assembly language concepts and abstract further.

Abstractions need to cover the computational behaviours (abstracting features of the physical processes in $\widehat{\mathbf{H}}$ to become computational features of the eventual DSL; programs written in the DSL are the $C$ of fig.\ref{fig:compcycle}).
They also need to cover the input and output mappings (the mappings $\widehat{\mathcal{R}}$ and $\mathcal{R}$ of fig.\ref{fig:compcycle} respectively):
how are abstract inputs to be implemented physically,
and how are physical outputs to be interpreted abstractly?

\subsubsection{Example}
In our example system, we draw on concepts from classical boolean circuits, and the standard quantum computing  gate model.

\paragraph{Behaviours.}
We have a physical interferometer comprising an arrangement of multiple beam splitters and phase shifters, each with potentially different properties. Collections of photons are fed in to different poistions in the interferometer, interact in the beam splitters and are modified in the phase shifters, and output photons are detected.
We capture the concept of components as Gates,
and how they modify photons as their GateBehaviour.

\paragraph{Inputs.}
There are distinct kinds of input.
First, there is the arrangement of the interferometer structure, of how the components are connected; this is a relatively fixed hardware configuration.
We capture this concept of arrangement as a fixed Circuit
and of the components as Gates placed on fixed Modes (GatePosn) in the Circuit.

Second, there are the values of the component parameters (transmission coefficients, phase changes, loss rates), which can be programmatically changed.
We capture this concept of parameter values as configuration parameters of a gate, as GateConfig,
combined as a CircuitConfig.
This configuration is set for a particular Run (a set of photons passed through the interferometer),
but may be changed between Runs.

Third, there are the input photons themselves:
the number of photons presented in each input mode for a given Run. 
At this point it is not clear how problem data might be encoded in lists of photons,
and this might be problem-specific.
For example, a problem input might be naturally encoded in a list of photons,
or it might need some `dual-rail' encoding (a zero encoded as [0,1], a 1 encoded as [1,0]) to ensure the number of photons does not vary wildly with abstract input value,
or some other encoding might be appropriate.
So we leave the input encoding at the primitive level of abstract mappings from the data to number of photons in individual modes.
Then the programmer is responsible for encoding their problem values into a suitable list of numbers.

\paragraph{Outputs.}
The output from an interferometer run is an observed pattern of photons in the output modes.
As with the input encoding, we use the straightforward representation that the physical photons map to abstract lists of numbers.

%................................
\subsection{Define a Domain Specific Language}\label{sec:method:dsl}

Use the abstract metamodel concepts to define a DSL \cite{Fowler2011}, both its syntax and semantics, that can be used to write programs that target the defined physical system.
This language should provide support for the concepts, and also respect any physical limitations of the device.

\subsubsection{Example}

We define a simple \textit{internal DSL} (sec.\ref{sec:impl_DSL}), designed for use within a Python framework that supports the DSL's semantics (through appropriate use of the simulator).

\paragraph{Syntax.}
The language defines the structure of a Circuit and its inputs, and the result of an execution. 
We define an abstract syntax and a concrete syntax.

An abstract syntax does not specify how the various terms are written in textual form.
This reduces clutter from including keywords, commas, brackets, etc., enabling the structure to be more visible.
The abstract syntax of our DSL, in Backus--Naur form (BNF) is:
\vspace{3mm} % bnf env does not leave any vertical space

\begin{bnf}
<ProbFn> ::= <In> 
    <CircuitPosn> <CircuitConfig> 
;;
<In> ::= $\mathbb{N}^+$ 
;;
<CircuitPosn> ::= $\mathbb{N}$ <GatePosn>$^+$
;;
<GatePosn> ::= <GateType> $\mathbb{N}^+$ 
;;
<GateType> ::= `P' // `MG' // `MGL1' // `MGL2' 
;;
<CircuitConfig> ::= <GateConfig>$^+$
;;
<GateConfig> ::=  <GateType> $\mathbb{R}^+$ 
;;
\end{bnf}
\vspace{3mm} % bnf env does not leave any vertical space

\noindent
In is a list of photon numbers that are the input to the various modes of the overall Circuit.
CircuitPosn is the number of modes, and a list of gate positions;
a gate position includes the gate type, and the modes it is applied to.
CircuitConfig is a list of gate configurations;
a gate configuration includes the gate type, and a list of relevant parameters.

A concrete syntax is a definition of the textual form of the abstract syntax: how the program is written down.
Different concrete syntaxes can be defined for different purposes,
such as for an in-program structure,
or for a configuration file.
Since we are defining an internal DSL for use with Python,
we define a concrete syntax in terms of Python dictionaries, for the contents of the various syntactical components,
and Python function calls, for defining programs using these components.
For example, the dictionary defining a mixer gate 
with parameter values $[\pi/4, 2\pi/3]$ is:
\begin{lstlisting}
my_mg_dict = {
    'name'  : 'MG',     # mixer gate
    'theta' : pi/4,     # angle
    'phi'   : 2*pi/3    # phase shift
}
\end{lstlisting}

We define some syntax checking, to define additional constraints beyond correct syntax of dictionaries and function calls.
For example, we check that the necessary keys are present in the relevant dictionaries,
and that their values have the required type ($\mathbb{N}$ or $\mathbb{R}$).

\paragraph{Semantics.}
The semantics defines the meaning of executing a DSL program.
There are many approaches to defining programming language semantics; we take a denotational approach \cite{Allison1986}.
This approach defines the meanings of language components as what they denote in mathematical terms;
these are composed to form the meaning of higher level language components.

We first define a static semantics, where the meaning of a component is True if it and its sub-components are well-formed, else it is False.
The static semantics ensures several consistency properties,
such as that input lists have the same length as the number of modes, that two-mode gates have two distinct mode positions, that CircuitPosn and CircuitConfig lists have the same gate types at the same locations,
and that photon number is conserved in the non-lossy gate case.

The dynamic semantics is the usual execution meaning, here written $\mathcal{M}(\mbox{X})$ for the meaning of a term X.
Here the meaning definition is heavily based on the mathematical model of the corresponding physical circuit.
We say that the dynamic semantics is \textit{undefined} if the static semantics is False.
This simplifies definitions by not having to specify a well-formedness precondition: that check has been covered in the static semantics.

In summary:

\begin{itemize}
    \item $\mathcal{M}$(GateConfig) is a $1\times 1$ (phase gate), $2\times 2$ (mixer gate), or $4\times 4$ (lossy mixer gate) matrix encoding the relevant configuration values. For example,
    \begin{equation}
            \mathcal{M}( \mbox{GateConfig}(\mbox{`MG'},[ \theta, \phi]) ) :=  
    \begin{pmatrix}
        t & r\\
        -r^{*} & t
    \end{pmatrix}
    \end{equation}
    where $t = \cos \theta$ is the transmission amplitude and $r = e^{- i \phi} \sin \theta$ is the reflection amplitude with a phase shift.
    \item $\mathcal{M}$(CircuitPosn, CircuitConfig) is an NMode $\times$ NMode matrix made from the smaller GateConfig matrices, placed on modes according to their Gate\-Posn.
    \item $\mathcal{M}$(ProbFn) is the probability mass function over all output configurations that can result from the input In applied to $\mathcal{M}$(CircuitPosn, CircuitConfig).
    Here, that is the result specified by the physical model definition (sec.\ref{sec:eg:model}).
    \begin{equation}
        \mathcal{M}\mbox{(ProbFn(In, CircuitPosn, CircuitConfig))  :=  model definition }
    \end{equation}
\end{itemize}
Programs can be written using standard Python constructs to compose multiple calls to PyBos functions with suitably arranged parameters along with  calls to other Python functions to implement other parts of algorithms; see sec.\ref{sec:method:simulate}.  

%..................................
\subsection{Build a reference implementation}

Having a language definition, the next step is to build a reference implementation (see sec.\ref{sec:impl_DSL}).
For an external (stand-alone) DSL, this will require building a compiler that implements the semantics for a software implementation of DSL programs.  
For an internal DSL, as here, this will require implementing a framework, and possibly a parser to convert a program into framework function calls, for a software implementation.
Both cases require support to define configurations and inputs for the physical device, to target hardware implementations.

\subsubsection{Example}
We have built a reference implementation of our DSL, called PyBos.
The physical system simulator (sec.\ref{sec:method_build_sim}) forms the basis of the framework for software implementation.
The function definitions are structured to mirror the DSL syntax,
so there are specific functions that implement the (dynamic semantics of) gates, circuits composed from gates, and full runs, along with many utility functions.
For example, the PyBos code to create a mixer gate matrix with parameter values $[\pi/4, 2\pi/3]$, as in $\mathcal{M}$(GateConfig(`MG',[$\pi/4$, $2\pi/3$])), is:
\begin{lstlisting}
import numpy as np
from pybos_sampler.core.gates import get_gate_mixer as MG
# create the mixer gate
my_mg = MG(theta=np.pi/4, phi=2*np.pi/3)
print(my_mg)
# prints the matrix:
# array([[ 0.70710678+0.j        , -0.35355339-0.61237244j],
#        [ 0.35355339-0.61237244j,  0.70710678+0.j        ]])
\end{lstlisting}

%..................................
\subsection{Implement programs in simulation}\label{sec:method:simulate}

The DSL and its reference implementation should first be validated by developing some small reference programs that between them exercise the entire language and implementation.
Is the language sufficiently expressive to program the desired systems?
Does the reference implementation support the full DSL correctly?
Does the implementation give correct results?
How does its performance scale with problem size?
What are the engineering limitations on device size and capacity, and therefore what is the maximum expected performance?
Some iteration may be necessary to enhance the DSL and its implementation,
but care should be taken to ensure that physical constraints are not violated at this stage.

\subsubsection{Example}
We have programmed and implemented (in simulation, for small systems) two non-lossy case studies.
Pseudocode for the original inspiration of Boson Sampling is  a single call to the top level language construct:

\begin{algorithm}[H]
\vspace{1mm}
\SetKwProg{Def}{def}{}{}
\SetKwInOut{Input}{input}
\SetKwData{In}{in}
\SetKwData{CircuitConfig}{circuitConfig}
\SetKwData{CircuitPosn}{circuitPosn}
\SetKwFunction{Pdf}{ProbFn}
\SetKwFunction{BosonSample}{BosonSampler}
    %\caption{Boson Sampling in PyBos pseudocode}\label{algo:M_BS}
    \Def{\BosonSample{\In, \CircuitPosn, \CircuitConfig}}{
    \Input{\In:list[$\mathbb{N}$], \CircuitPosn:dict, \CircuitConfig:dict}
    %\BlankLine 
    $pf$ := \Pdf{\In, \CircuitPosn, \CircuitConfig}
    \\
    \Return $pf$
    }
    \vspace{2mm}
\end{algorithm}

Pseudocode for an optimisation problem that learns a set of circuit configuration parameters that, given a pattern of input photons and gate positions, produces some target probability mass function (pmf) is:

\begin{algorithm}[H]
\vspace{1mm}
\SetKwProg{Def}{def}{}{}
\SetKwInOut{Input}{input}
\SetKwData{In}{in}
\SetKwData{config}{config}
\SetKwData{target}{target}
\SetKwData{Ntrain}{NTrain}
\SetKwData{CircuitPosn}{circuitPosn}
\SetKwFunction{Pdf}{ProbFn}
\SetKwFunction{Optimiser}{optconfig}
    %\caption{Configuration Optimisation in PyBos pseudocode}\label{algo:M_Config_opt}
    \Def{\Optimiser{\In, \CircuitPosn, \target}}{
        \Input{\In:list[$\mathbb{N}$], \CircuitPosn:dict, \target:list[prob], \Ntrain:$\mathbb{N}$}
        \BlankLine 
        \textit{config} := initial random gate configurations
        \\
        \For(){n in $1$ to \Ntrain}{
            \textit{pf} := \Pdf\!(\In, \CircuitPosn, \textit{config})
            \\
            \textit{config} := update(\textit{config}, \textit{pf}, \target)
        }
        \BlankLine
    \Return \textit{config}
    }
    \vspace{1mm}
\end{algorithm}

\noindent
Here there are multiple calls to the top level language construct,
interleaved with calls to a classical training algorithm.
The `update' function could be any suitable classically computed machine learning function
\cite{Park:2025-UCNC}.
The result is a circuit configuration that implements the required pmf.
An analogous algorithm can be written to optimise gate numbers and positions for the same problem.

An extension is to optimise a single system to output different target pmfs for different inputs: a classifier.
One aim of the overall project is to determine if and how loss can be exploited to improve performance of such optimisation algorithms,
by selectively losing the unwanted information.

%..................................
\subsection{Implement programs physically}

The DSL and its hardware support should be validated by porting the reference programs from sec.\ref{sec:method:simulate} to a physical implementation.
Does the implementation correctly specify the hardware configuration?
Can the hardware be configured dynamically as necessary?
Can the inputs be provided correctly, and the outputs obtained correctly?
Does the physical implementation give the correct results (to within an acceptable tolerance)?
If not, is the problem with the physical model, the abstractions in the metamodel and DSL, the DSL implementation, the program, the hardware device driver, or elsewhere?
Does the physical implementation exhibit the expected performance?
How does its performance scale with problem size?
What are the engineering limitations on device size and capacity, and therefore what is the maximum expected performance?

\subsubsection{Example}
We have not yet reached this step in our current project, although we have been liaising with the experimentalists to confirm several features of the physical system.

When we reach this step, our initial focus will be on checking that the model of the lossy gates is adequate for the physical implementation,
then that the semantics definitions of higher level components correctly captures the physical behaviours,
then an implementation of the reference programs.

In particular, both the simulation and the physical implementation are only approximations to the actual probability mass function.
The pmf output from a \textit{simulation} execution is 
typically approximated by thresholding to make the run-time feasible;
better approximations can be made by decreasing the threshold.
The output from a \textit{physical} execution (itself comprising multiple applications of the input and measuring of outputs) is an experimentally derived approximation of the pmf;
better approximations can be made by increasing the number of input applications.
A part of the validation will be to ensure that these approximations are adequate to the task, and to determine the appropriate precisions needed.

%=========================
\section{Discussion and Conclusion}

We have presented a methodology for designing models of computation,
and associated domain specific programming languages,
that can exploit the exotic computational properties of novel physical devices.
The example we have provided defines a DSL
that is a low-level language for programming bosonic interferometer devices to perform computation tasks,
and illustrated how this might be used for machine learning tasks.

We expect that application of the methodology will yield such low-level languages in the first instance.
This is only to be expected, and mirrors the development of classical digital computing.
Once such unconventional `assembly languages' have been studied,
higher level abstractions enabling more structured application to more sophisticated problems,
should be forthcoming.

In future work, we plan to examine how our approach can be used to exploit the existence of loss in such physical devices as a feature, rather than a bug.

%\begin{credits}
\subsubsection*{Acknowledgements} This work was funded
by the LoCoMo (Lossy Computational Models) project,
ARIA `Nature Computes Better', grant number 20240730 NACB-SE02-P02.

% \textbf{\discintname}
% The authors have no competing interests to declare that are
% relevant to the content of this article. 
%\end{credits}
\bibliographystyle{plainnat}
\bibliography{main}
\end{document}